\documentclass[aps,prl,twocolumn,floatfix,showpacs,superscriptaddress]{revtex4}
\usepackage{graphicx}
\usepackage{bbm}
\usepackage{amsmath,amssymb}
\newcommand{\be}{\begin{equation}}
\newcommand{\beq}{\begin{equation}}
\newcommand{\ee}{\end{equation}}
\newcommand{\bea}{\begin{eqnarray}}
\newcommand{\eea}{\end{eqnarray}}
\newcommand{\ba}{\begin{array}}
\newcommand{\ea}{\end{array}}

\begin{document}
\title{Electron-Electron Interactions in Artificial Graphene}

\author{E. R\"as\"anen}
\email[Electronic address:\;]{erasanen@jyu.fi}
\affiliation{Nanoscience Center, Department of Physics, University of
  Jyv\"askyl\"a, FI-40014 Jyv\"askyl\"a, Finland}
\affiliation{Physics Department, Harvard University, Cambridge, 
Massachusetts 02138, USA}
\author{C. A. Rozzi}
\email[Electronic address:\;]{carloandrea.rozzi@nano.cnr.it}
\affiliation{CNR - Istituto Nanoscienze, Centro S3, 
via Campi 213a, I-41125 Modena, Italy}
\author{S. Pittalis}
\affiliation{Department of Physics and Astronomy, University of Missouri, Columbia, Missouri 65211, USA}
\affiliation{Department of Chemistry, University of California, Irvine, California 92697, USA}
\author{G. Vignale}
\affiliation{Department of Physics and Astronomy, University of Missouri, Columbia, Missouri 65211, USA}


\begin{abstract}
Recent advances in the creation and modulation of graphene-like systems are
introducing a science of ``designer Dirac materials''. 
In its original definition, artificial graphene is a man-made nanostructure 
that consists of identical potential wells (quantum dots) arranged in a 
adjustable honeycomb lattice in the two-dimensional electron gas. 
As our ability to control 
the quality of artificial graphene samples improves, so grows the need 
for an accurate theory of its electronic properties, including the effects 
of electron-electron interactions. Here we determine those effects 
on the band structure and on the emergence of Dirac points.
\end{abstract}

 
\maketitle

Graphene, a single two-dimensional (2D) crystal of carbon 
atoms, has become the most attractive carbon-based
material and one of the hottest topics in condensed
matter and material physics~\cite{geimRMP}.
Many of the superlatives attributed to graphene are
due to the fact that, at low energies, electrons (and holes)
behave as massless chiral Dirac fermions as a result of
a linear dispersion relation near two inequivalent corners of
the Brillouin zone, i.e., the Dirac points.

Not surprisingly, material combinations and arrangements 
with properties similar to real graphene have been 
sought~\cite{esslinger,Manoharan}, and experimental progress
in producing artificial structures has been impressive.
In 2009, the use of a hexagonal nanopatterned 2D electron 
gas (2DEG) -- similar to previously considered triangular 
antidot arrays~\cite{flindt,flindt2} -- was suggested~\cite{parklouie},
and soon thereafter Gibertini {\em et al.}~\cite{gibertini} presented the 
graphene-like band structure of GaAs quantum dots (QDs) arranged in a 
honeycomb lattice -- a system they called as
``artificial graphene''  (AG). Their tight-binding calculations
were supplemented by an experimental demonstration
of a nanopatterned modulation-doped sample~\cite{gibertini,simoni}.
Very recently, the same authors and their
collaborators showed that AG subjected to 
a strong magnetic field exhibits collective modes according to the 
Mott-Hubbard model~\cite{science}.
In a parallel development, a structure equivalent to AG has  been fabricated 
by STM-controlled deposition of CO molecules on the 2DEG on Cu (111) surface~\cite{Manoharan}. 
This leads to extremely controlled samples having even less defects than
natural graphene. The trend limited to electronic systems:
as another way to control Dirac fermions Tarruell 
{\em et al.}~\cite{esslinger} have created a tunable honeycomb optical 
lattice in an ultracold quantum gas.

One of the reasons for pursuing the study of AG
is that this system offers the opportunity to experimentally study
regimes that are difficult or impossible to achieve in natural graphene,  
(including high magnetic fluxes,  tunable lattice constants, 
precise manipulation of defects, edges, and strain) and
allow the experimental observation 
of several predictions made for massless Dirac 
fermions~\cite{Barlas,Hwang,PoliniAsgari}.
In Ref.~\cite{karel} 
experimental criteria for the realization of graphene-like physics 
in 2DEG have been described. 
Recently, the system has been proposed as a candidate for the 
observation of a quantized anomalous Hall insulator~\cite{zhang}.

These rapid experimental advances call for state-of-the-art 
numerical tools to calculate electronic properties of AG 
and, more generally, of artificial lattices.  In view of the 
accuracy with which the sample can be prepared, it is particularly 
important to be able to reliably predict the conditions under which 
isolated Dirac points will appear in AG. 
By ``isolated Dirac points'' we mean a set of points in momentum space 
where a conical conduction band makes contact with a conical valence 
band -- the contact occurring at an energy at which no other state exists. 
As shown in Ref.~\cite{gibertini} such points occur only if the
depth of the potential within the QDs of the AG structure exceeds a 
certain minimum value.  However, this minimum value cannot be reliably 
predicted from a theory that neglects the electron-electron (e-e) 
interactions.  
It is not just a matter of replacing the bare potential by an effective 
one that includes interaction effects such as screening, exchange and 
correlation.  The key point is that this effective potential must be 
consistent with an electronic density distribution that places the Fermi 
level at the Dirac point.  For example, the effective potential at $N=1$ 
electrons per dot is quite different from the effective potential at 
$N=4$ electrons per dot, because the electronic density distributions 
in the two cases are widely different.

In order to include the e-e interactions in the study of 
AG structures we resort to density-functional  
theory~\cite{dft} (DFT) within the 2D version of the 
local-density approximation (2D-LDA) that has been shown to successfully
describe the electronic structure of individual and coupled QDs  
fabricated in the 2DEG~\cite{reimann,other}.
This takes us two steps beyond
previous tight-binding studies.  First, we are  able to 
include the e-e interactions, thus producing the 
first self-consistent (density-dependent) band structure of AG. 
Second, we are also able to describe the system in a fully resolved 
manner in real space with a realistic model potential. 
Our results fully confirm the existence of Dirac points
and massless Dirac fermions. However, we find that the 
threshold for the emergence of isolated Dirac points is 
increased in a way that significantly depends on the electron density.  
For example, from the non-interacting electron theory~\cite{gibertini} 
we know that both $N=1$ and $N=4$ electrons per dot are candidates 
for the emergence of an isolated Dirac point, even though the $N=4$ 
case requires a much deeper potential well. When e-e 
interactions are included we find that the threshold for the emergence 
of the Dirac point is  slightly enhanced in the $N=1$ case, but largely 
enhanced in the $N=4$ case. 

\begin{figure}
\includegraphics[width=0.99\columnwidth]{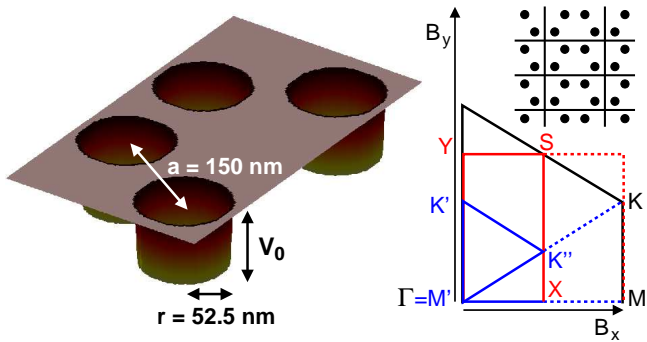}
\caption{(Color online) 
Left: Rectangular unit cell in real space.
Right: Orthorhombic irreducible Brillouin zone (rectangle $\Gamma X S Y$) 
compared with the conventional one ($\Gamma M K$) for the honeycomb lattice.
The basis vectors ${\bf B}_x$ and ${\bf B}_y$ are defined in the text.
 The hexagonal high symmetry points are 
${\bf K}\equiv\frac{1}{3}{\bf B}_y={\bf K^\prime}$, ${\bf M}\equiv\boldsymbol{\Gamma}={\bf M^\prime}$.  
The mapping from the hexagonal to the orthorombic cell is performed as follows: the $\Gamma-M$ line corresponds to the $\Gamma-X-\Gamma$ 
path in the orthorhombic cell; the $M-K$ segment corresponds to $\Gamma-K^\prime$, and the $K-\Gamma$ 
segment corresponds to the two segment path $K^\prime-K^{\prime\prime}-M^\prime$, with ${\bf K^{\prime\prime}}\equiv\frac{1}{2}\left({\bf B}_x+\frac{1}{3}{\bf B}_y\right)$. 
Inset: part of the lattice in real space. }
\label{fig1}
\end{figure}

We consider electrons confined in GaAs/AlGaAs 
QDs in the effective mass approximation,
which is the conventional approach to the
modeling of individual and coupled semiconductor
QDs~\cite{reimann,kouwenhoven}. We use the effective mass
$m^*=0.067\,m_0$ and the dielectric constant
$\epsilon=12.4\,\epsilon_0$. Each QD is
modeled by a cylindrical hard-wall potential
with radius $r=52.5$ nm and a 
tunable height $V_0$, and the lattice
constant is fixed to $a=150$ nm (see Fig.~\ref{fig1}). 
These are values that can be reached 
experimentally~\cite{Hirjibehedin02,Hirjibehedin05,Garcia05} 
and they are
similar to the previous tight-binding study in 
Ref.~\cite{gibertini} in order to allow 
direct comparison between the results.
We point out that softening the potential from 
a hard-wall to a Gaussian shape did 
not have qualitative effects on the results below.

The electronic structure of the system was calculated with the Octopus package~\cite{octopus}. 
The code solves the Kohn-Sham equations on a real-space discrete grid. The only required 
convergence parameter is the grid spacing, and it is therefore not bound to any particular 
choice of a basis set. 
In order to describe a truly 2D distribution of atoms we follow the procedure described in Ref.~\cite{cutoff}, i.e., we impose a  set of mixed boundary conditions (periodic in the plane, zero Dirichlet off-plane) and  cut-off the Coulomb potential  to zero along the direction perpendicular to the plane, while retaining its full long range of action within the plane. The levels obtained are rigorously equivalent to those calculated in an infinitely wide supercell in the direction perpendicular to the plane. 

As shown in Fig.~\ref{fig1},  the unit cell is chosen 
not to be minimal but to contain four dots, so that the 2D Bravais lattice becomes rectangular.  The unit cell size is $\sim 130\,{\rm nm}\times\,225\,{\rm nm}$  
and the
grid spacing is $\sim 2.45\,{\rm nm}$. The reciprocal space cell is generated by the vectors 
${\bf B}_x = \frac{2\pi}{a}\left[\frac{1}{3},0\right]$ and ${\bf B}_y = \frac{2\pi}{a}\left[0,\frac{1}{\sqrt{3}}\right]$,  where $a$ is the interdot distance. The volume of this cell in reciprocal space is half the 
volume of the standard hexagonal BZ. Obviously, the physically meaningful results are 
not affected by our choice of the unit cell. However, in order to compare to calculations 
performed in the minimal (standard) cell, our bands must be appropriately unfolded. 
The high symmetry lines of the standard BZ can all be mapped to corresponding paths into the 
smaller BZ.  The mapping is described in the caption of Fig.~\ref{fig1}.  For ease of  comparison with previously published results, all the bands are displayed unfolded 
in this work. In the process of unfolding the band structure from the rectangular to the hexagonal cell, care must be exerted to avoid the phenomenon of {\it aliasing}, i.e., the spurious duplication of bands.  This is greatly facilitated by the observation that the true bands must be continuous and differentiable at $K$ along the $\Gamma-K-M$ line.

As mentioned in the introduction, in order to assess 
the importance of e-e interactions, we compare the results for noninteracting
electrons with those computed using the Kohn-Sham
DFT approach~\cite{dft} within the 2D LDA.
For the correlation part of the LDA we have used the
parametrized form of the quantum Monte Carlo data 
calculated for the 2DEG by Attaccalite
{\em et al.}~\cite{attaccalite} In view of previous
works on QDs fabricated in the 2DEG~\cite{reimann,other} 
we believe that the 2D-LDA provides a reasonable approximation
for the energy bands considered here. 

\begin{figure}
\includegraphics[width=0.99\columnwidth]{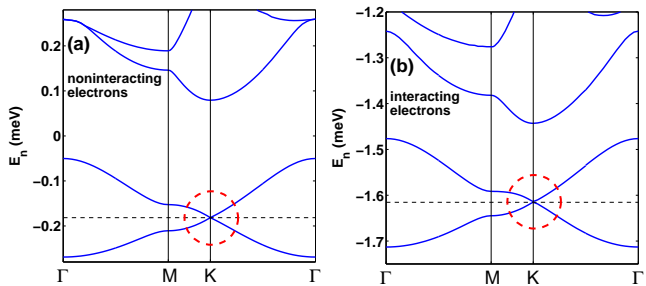}
\caption{(Color online) Energy bands calculated for
noninteracting (a) and interacting electrons ($N=1$ per dot) 
in density-functional theory (b). Clear Dirac points
and a linear dispersion relation can be found in
both cases (circles).
}
\label{bandfigure}
\end{figure}

In Fig.~\ref{bandfigure} we show the energy bands
calculated for noninteracting (a) and interacting
electrons in the LDA (b) when $V_0=-0.6$ meV. 
In both cases we find distinctive Dirac points at K
with a linear dispersion relation. These are the
defining attributes of graphene-like physics.
From Fig.~\ref{bandfigure} it is clear that, in the 
case of singly occupied QDs in the AG, the e-e interactions
do not make the Dirac point less stable.
In general, the band structures are very similar,
although, as expected, the e-e interactions
reduce the bandwidth and thus increase the energy
gaps between the two lowest bands. However, as an interesting
exception to this tendency, the LDA result
has a considerably smaller gap above the
two lowest bands. 
We believe that this indicates the gap has a primarily kinetic
origin, i.e. it arises from an overlap of neighboring localized orbitals
forming a bonding-antibonding pair. Such a gap is reduced when
the overlap of orbitals in the pair decreases, due to increased 
electron localization.

\begin{figure}
\includegraphics[width=0.90\columnwidth]{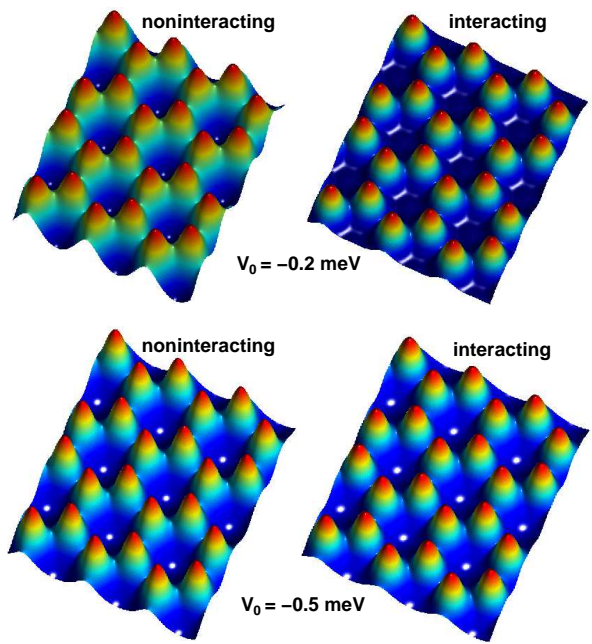}
\caption{(Color online) Electron densities in artificial
graphene ($N=1$ per dot) calculated for two depths of the potential
wells, $V_0=-0.2$ meV (up) and $V_0=-0.5$ meV (down).
The left and right panels show the results without
and with the e-e interactions, respectively.
}
\label{density}
\end{figure}

The general tendency of increased electron localization 
due to e-e interactions is clearly visible in the electron density
shown in Fig.~\ref{density}.
As expected,  the relative difference in the density
with and without e-e interactions is larger when $|V_0|$ is
small.  Indeed, for small $|V_0|$  the system is closer to 
the homogeneous 2DEG,
where interactions (at small densities as in this case)
drive the system to the Wigner crystal regime~\cite{vignalebook}.

\begin{figure}
\includegraphics[width=0.95\columnwidth]{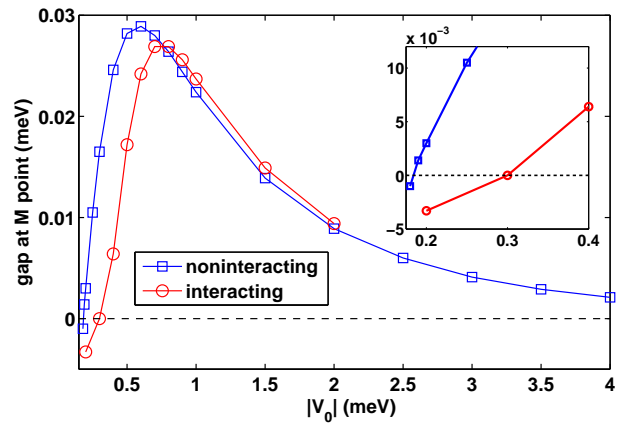}
\caption{(Color online) Energy gap at the M point
in the Brillouin zone as a function of the quantum-dot
depth. Results without (squares) and with e-e interactions (circles)
are shown for $N=1$ per dot. 
The inset shows a zoom of the regime where the gap opens.
}
\label{gap}
\end{figure}

To obtain a closer view on the stability of the Dirac points and
the effects of e-e interactions,
we next examine the onset and the size of the energy gap 
at the M point (see Fig.~\ref{bandfigure}). The gap is
defined as the energy difference between the Fermi level
and the band closest to the Fermi level at the M point. Negative
values correspond to crossings of the band(s) through the Fermi 
level, so that then there is no gap around the Dirac point.
Figure~\ref{gap} shows the size of the gap as
a function of $|V_0|$, i.e., the depth of the QDs. 
In the absence of interactions the threshold potential for the 
gap is $V_0=-0.18$ meV which agrees perfectly with the tight-binding 
result in Ref.~\cite{gibertini}. When the interactions
are included in the LDA level, the threshold shifts to
$V_0=-0.30$ meV.  In other words, the inclusion of e-e 
interactions leads us to predict that  
a deeper QDs is required in order to achieve an isolated Dirac 
point in the system.  
When $|V_0|$ is further increased  the
results without and with interactions become very similar.
The gap reaches its maximum value at
$V_0=-0.8$ meV  (or $V_0=-0.6$ meV in the absence of interactions),
which can be regarded as the optimal 
QD depth for the stability of the Dirac point.
As expected, the gap shown in  Fig.~\ref{gap} closes asymptotically
in the large-$|V_0|$ limit. The closing proceeds 
in a similar manner regardless of
e-e interactions -- albeit LDA results are not available at $|V_0|>2$ meV
due to poor convergence. Gibertini {\em et al.} argued that
there is a localization threshold in the system at  
$|V_0|\sim 3$ meV, due to the formation of disorder-induced 
bound states. This effect is
not included in our calculations, 
since the perfect periodicity of the lattice
keeps the wave functions extended, in principle up to
the limit $|V_0| \rightarrow \infty$.

\begin{figure}
\includegraphics[width=0.99\columnwidth]{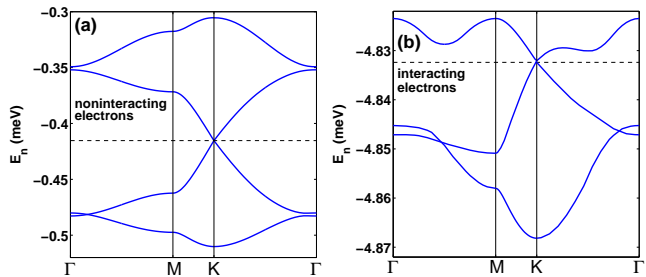}
\caption{(Color online) Energy bands close to
the Fermi energy calculated for
noninteracting (a) and interacting electrons, when
there are four electrons in each quantum dot.
The two lowest bands (not shown) are located 
at about $-1.3$ and $-5.6$ meV in (a) and (b),
respectively.
}
\label{fig5}
\end{figure}

Finally we consider occupying the QDs with more than one electron.
As suggested in Ref.~\cite{gibertini}, where a noninteracting
tight-binding scheme was used, the next Dirac point
should appear when $N=4$. In Fig.~\ref{fig5} we show the 
energy bands close to the Fermi energy when $N=4$ and
$V_0=-2$ and -6.8 meV in the noninteracting (a) and interacting
cases (b), respectively. The lowest two bands (not shown) 
are located at considerably lower energies, i.e., 
at about $-1.3$ and $-5.6$ meV, respectively.
It is noteworthy that although the Dirac point can be found
in both cases, it is much less stable in the interacting system:
first, due to the strong {\em intradot} interactions the QDs need 
to be significantly deeper than without interactions. Furthermore,
the Dirac point in the interacting case appears in an energy gap 
that is smaller by about an order of magnitude. Consequently,
decreasing the potential depth from $|V_0|=6.8$ meV to
 6.5 meV already disruptes the Dirac point, as the band above $E_F$
in Fig.~\ref{fig5}(b) is shifted downwards. In contrast, the Dirac point in the 
noninteracting case is stable down to about $|V_0|=1$ meV.
These results indicate that the realization of AG constitutes a 
challenge for the present engineering techniques if the QDs are occupied 
by several electrons.

To summarize, we have studied the electronic properties of
artificial graphene fabricated in the two-dimensional electron gas.
The electron-electron interactions,
treated here within the two-dimensional 
local-density approximation in density-functional 
theory, generally lead to stronger localization of the electrons in 
the quantum dots.  The interactions shift the threshold for the
emergence of isolated Dirac points to larger well depths than
found without interactions. This effect is significantly pronounced
when the number of electrons is increased. This sets a particular
challenge for the realization of Dirac points 
if the quantum dots in the hexagonal lattice are occupied
by more than a single electron.

\begin{acknowledgments}
This work has been  supported by 
the Academy of Finland, the Wihuri Foundation, 
and the DOE grant DE-FG02-05ER46203 (SP).
CSC Scientific Computing Ltd.
is acknowledged for computational resources.   
\end{acknowledgments}

\end{document}